# ATOMS: ALMA Three-millimeter Observations of Massive Star-forming regions -IV. Radio Recombination Lines and evolution of star formation efficiencies

C. Zhang,[1,2]⋆ Neal J. Evans II,[3] T. Liu,[4]† J.-W. Wu,[5,6] Ke Wang,[7] H.-L. Liu,[8] F.-Y. Zhu,[4] Z.-Y. Ren,[6]
L. K. Dewangan,[9] Chang Won Lee,[10,11] Shanghuo Li,[10] L. Bronfman,[12] A. Tej,[13] D. Li,[6,14]

[1]*Department of Physics, Taiyuan Normal University, Jinzhong 030619,China*
[2]*Institute of Computational and Applied Physics, Taiyuan Normal University, Jinzhong 030619,China*
[3]*Department of Astronomy, The University of Texas at Austin, 2515 Speedway, Stop C1400, Austin, Texas 78712-1205, U.S.A.*
[4]*Shanghai Astronomical Observatory, Chinese Academy of Sciences, 80 Nandan Road, Shanghai 200030, People's Republic of China*
[5]*University of Chinese Academy of Sciences, Beijing 100049, PR China*
[6]*National Astronomical Observatories, Chinese Academy of Sciences, Beijing, 100012, People's Republic of China*
[7]*Kavli Institute for Astronomy and Astrophysics, Peking University, 5 Yiheyuan Road, Haidian District, Beijing 100871, People's Republic of China*
[8]*Department of Astronomy, Yunnan University, Kunming,650091, PR China*
[9]*Physical Research Laboratory, Navrangpura, Ahmedabad–380 009, India*
[10]*Korea Astronomy and Space Science Institute, 776 Daedeokdaero, Yuseong-gu, Daejeon 34055, Republic of Korea*
[11]*University of Science and Technology, Korea, 217 Gajeong-ro, Yuseong-gu, Daejeon 34113, Republic of Korea*
[12]*Departamento de Astronomía, Universidad de Chile, Casilla 36-D, Santiago, Chile*
[13]*Indian Institute of Space Science and Technology, Thiruvananthapuram 695 547, Kerala, India*
[14]*NAOC-UKZN Computational Astrophysics Centre, University of KwaZulu-Natal, Durban 4000, South Africa*

**ABSTRACT**

We report detection of radio recombination line (RRL) H$_{40\alpha}$ toward 75 sources, with data obtained from ACA observations in the ATOMS survey of 146 active Galactic star forming regions. We calculated ionized gas mass and star formation rate with H$_{40\alpha}$ line emission. The mass of ionized gas is significantly smaller than molecular gas mass, indicating that ionized gas is negligible in the star forming clumps of the ATOMS sample. The star formation rate (SFR$_{H_{40\alpha}}$) estimated with RRL H$_{40\alpha}$ agrees well with that (SFR$_{L_{bol}}$) calculated with the total bolometric luminosity (L$_{bol}$) when SFR $\geqslant$ 5 M$_\odot$ Myr$^{-1}$, suggesting that millimeter RRLs could well sample the upper part of the initial mass function (IMF) and thus be good tracers for SFR. We also study the relationships between L$_{bol}$ and the molecular line luminosities (L$'_{mol}$) of CS J=2-1 and HC$_3$N J=11-10 for all the 146 ATOMS sources. The L$_{bol}$-L$'_{mol}$ correlations of both the CS J=2-1 and HC$_3$N J=11-10 lines appear approximately linear and these transitions have success in predicting L$_{bol}$ similar to that of more commonly used transitions. The L$_{bol}$-to-L$'_{mol}$ ratios or SFR-to-mass ratios (star formation efficiency; SFE) do not change with galactocentric distances (R$_{GC}$). Sources with H$_{40\alpha}$ emission (or H II regions) show higher L$_{bol}$-to-L$'_{mol}$ than those without H$_{40\alpha}$ emission, which may be an evolutionary effect.

**Key words:** ISM: clouds, (ISM:) H II regions, stars: formation, radio lines: ISM

## 1 INTRODUCTION

Stars are known to form in molecular gas, but recent works suggest that dense gas, probed by species other than CO, provides a better predictor of the rate of star formation (Evans 2008; Vutisalchavakul et al. 2016). On the scale of galaxies, Gao & Solomon (2004) showed the luminosities of IR emission (L$_{IR}$ tracing the SFR) of entire galaxies have a tighter linear correlation with the HCN (J=1-0) than with the CO luminosity. This correlation even extends to Galactic dense clumps undergoing high mass star-formation (Wu et al. 2005). This so called "dense gas star formation law" suggests that the dense molecular gas rather than the total molecular gas is the direct fuel for star formation. Such linear correlations have also been found in other dense gas tracers (e.g. HCN J=4-3, HCO$^+$ J=4-3, CS J=7-6, CS J=2-1) with similar or higher critical densities (Wu et al. 2010; Lada et al. 2012; Zhang et al. 2014; Liu et al. 2016; Stephens et al. 2016; Tan et al. 2018; Liu et al. 2020b). Evans et al. (2014) compared various models of star formation to observations of the nearby clouds and found that the mass of dense gas was the best predictor of the star formation rate. Most recently, Vutisalchavakul et al. (2016) showed that a similar result applied to more distant and massive clouds in the Galactic Plane.

Accurate estimates of SFRs are crucial in studying the dense gas star formation law. Presently, a wide variety of SFR diagnostics are used in the literature (e.g., Kennicutt 1998; Kennicutt & Evans

⋆ E-mail: zhangchao920610@126.com
† E-mail: liutie@shao.ac.cn

2012), each of which has its strengths and shortcomings (Murphy et al. 2011). Among them, infrared or bolometric luminosity is commonly used in estimating SFRs in both external galaxies and Galactic clumps. Bolometric luminosity comes from heated dust grains surrounding massive star-forming regions and does not suffer significantly from extinction. While a portion of the bolometric luminosity will arise from dust heated by older stars (Hirashita et al. 2003; Bendo et al. 2010; Li et al. 2010) on galaxy scales, that is not significant on the scale of star formation regions in our Galaxy.

H II regions are hot ionized gas surrounding exciting central OB stars, and thus they are ideal tracers for new star formation in giant molecular clouds. The millimeter recombination lines originating from low energy levels (n=20–50) that would not be much affected by potential maser amplification or dust extinction, may be reliable and quantitative probes of the EUV continuum at 13.6 eV to above 54.6 eV (Scoville & Murchikova 2013). Strong correlation between the integrated 6 cm radio continuum and millimeter RRL (mm-RRL) emissions have been revealed in Galactic H II regions (Kim et al. 2017), indicating that mm-RRLs can trace the same ionized nebula as centimeter radio continuum. RRLs are therefore optimal tracers of H II regions and new star formation. The line luminosities of mm-RRLs are directly proportional to the emission measures of their ionized regions and thus can be related to the star formation rates (Scoville & Murchikova 2013; Bendo et al. 2015, 2016, 2017; Binder & Povich 2018a; Michiyama et al. 2020), if the initial mass function (IMF) is well sampled. In this paper, we test the use of $H_{40\alpha}$ RRL to trace UC H II regions and star formation rate in Galactic massive clumps.

We also investigate the "dense gas star formation law" with CS J=2-1 and $HC_3N$ J=11-10. Liu et al. (2020b) reported that the star formation scaling relationships of $H^{13}CN$ J=1-0 and $H^{13}CO^+$ J=1-0 lines as well as their corresponding main lines, with data obtained from ACA observations in the ATOMS survey, appear approximately linear. Table 1 lists the relevant parameters for the lines studied here and, for comparison, the lines studied by Liu et al. (2020b). CS J=2-1 and $HC_3N$ J=11-10 have higher effective excitation densities ($n_{eff}$) than HCN J=1-0 and $HCO^+$ J=1-0 (Shirley 2015), and similar to those of the rare isotopologues, $H^{13}CN$ and $H^{13}CO^+$. Therefore, they could be good tracers of more concentrated dense gas in molecular clouds. In contrast, the total luminosity of HCN and $HCO^+$ $J = 1 - 0$ lines can be dominated by low level emission from extended regions (Evans et al. 2020; Liu et al. 2020a). The $HC_3N$ J=11-10 line has an $n_{eff}$ about 10 times that of CS 2-1, and it is a good tracer of dense cores and filaments in Galactic clouds (Liu et al. 2020a). In addition, this line as well as other high J transitions of $HC_3N$ are also detectable in external galaxies (Lindberg et al. 2011; Aladro et al. 2011; Jiang et al. 2017), indicating that $HC_3N$ lines are potentially good tracers for studying the "dense gas star formation law" in both external galaxies and Galactic clouds.

The dispersion in the ratio of star formation rate to dense gas mass (or star formation efficiency; SFE) is still substantial. Multiple feedback mechanisms from formed OB stars can significantly reduce star formation rates and may play a major role in determining the star formation efficiency in molecular clouds (Krumholz et al. 2014). In this paper, we divide 146 ATOMS clumps into two groups, i.e., the one with and other without $H_{40\alpha}$ emission (or H II regions), and investigate how formed OB stars influence the star formation efficiency in these star forming clumps.

In section 2, we describe the observations. Section 3 presents the results, focusing on empirical relations between bolometric and line luminosities. We calculate star formation rates with RRLs, derive star formation efficiencies and gas depletion times, and provide a physical interpretation for the empirical relations in Section 4. We summarize our main conclusions in Section 5.

## 2 OBSERVATIONS

### 2.1 The sample

We made the ALMA observations for 146 active Galactic star forming regions as the ATOMS (ALMA Three-millimeter Observations of Massive Star forming regions) survey. These 146 sources were selected from the CS J = 2-1 survey of Bronfman et al. (1996), a complete and homogeneous molecular line survey of ultra compact (UC) H II region candidates in the Galactic plane. The sample of 146 targets is complete for proto-clusters with bright CS J = 2-1 emission ($T_b$ >2 K), indicative of reasonably dense gas. More parameters of 146 targets are shown in Table A of Liu et al. (2020a). Most (139) of the targets are located in the first and fourth Galactic Quadrants of the inner Galactic Plane. The distances of the sample clouds range from 0.4 kpc to 13.0 kpc with a mean value of 4.5 kpc. The sample includes 27 distant (d > 7 kpc) sources that are either close to the Galactic center or mini-starbursts, representing extreme environments for star formation. The properties of these sources have been described in detail in Liu et al. (2016, 2020a,b) and Liu et al. (2021).

### 2.2 ALMA ACA observations

The data on CS J=2-1, $HC_3N$ J=11-10 and $H_{40\alpha}$ RRL were obtained from the Atacama Compact 7 m Array (ACA; Morita Array) observations in the ATOMS survey. We use only ACA data to obtain a larger FOV than we would have with the 12-m array data and the ACA data can match clump-scale luminosity information and better trace the overall spatial distribution of gas within these star forming regions. The ACA observations were performed between September to mid November in 2019 in band 3 (Project ID: 2019.1.00685.S; PI: Tie Liu). The typical ACA observing time per source is ∼ 8 minutes. The angular resolution and maximum recovered angular scale for the ACA observations are ∼ 13.″1 - 13.″8 and ∼ 53.″8 - 76.″2, respectively. The CS J=2-1, $HC_3N$ J=11-10 and $H_{40\alpha}$ are included in three spectral windows in the upper sideband. The spectral resolution for CS J = 2-1, $HC_3N$ J=11-10 and $H_{40\alpha}$ is ∼1.5 km s$^{-1}$.

Calibration was carried out using CASA software package version 5.6 (McMullin et al. 2007). More details about ALMA observations can be found in Liu et al. (2020a,b).

## 3 RESULTS

### 3.1 Extraction of compact objects

We extracted compact objects from the integrated intensity maps of the three lines (CS J=2-1, $HC_3N$ J=11-10 and $H_{40\alpha}$) using an elliptical Gaussian fit. Figure 1 shows the integrated intensity maps for two example sources. The objects in molecular line emission can be easily identified by eye. In total, we detected 177, 185 and 75 sources in emission from CS J=2-1, $HC_3N$ J=11-10 and $H_{40\alpha}$, respectively. We also show the 3mm continuum emission for comparison, with parameters derived by Liu et al. (2020b), in the figure.

The majority (∼80%) of targeted sources contain only one object in molecular line emission as illustrated for I13134-6242 in Figure 1. We identified multiple objects in 28, 32 and 4 targeted sources from CS, $HC_3N$ and $H_{40\alpha}$ maps, respectively. In such sources, we separately fit every single object using an elliptical Gaussian fit. The parameters of every single object are summarized in Tables 2 to 4. When we calculate luminosity of the targeted source, we added all the multiple objects to obtain the overall clump properties.

**Table 1.** spectral lines (The E$_u$/k and effective excitation density are from Shirley 2015).

| Molecule | Transition | Rest frequency (GHz) | E$_u$/k (K) | velocity resolution (km/s) | n$_{eff}$(15K) (cm$^{-3}$) |
|---|---|---|---|---|---|
| CS | 2-1 | 97.980953 | 7.05 | 1.5 | 1.5×10$^4$ |
| HC$_3$N | 11-10 | 100.07639 | 28.82 | 1.5 | 1.6×10$^5$ |
| H$_\alpha$ | H$_{40\alpha}$ | 99.022952 | | 1.5 | |
| HCN | 1-0 | 88.631847 | 4.25 | 0.2 | 5.6×10$^3$ |
| HCO$^+$ | 1-0 | 89.188526 | 4.28 | 0.2 | 6.4×10$^2$ |
| H$^{13}$CN | 1-0 | 86.339918 | 4.14 | 0.4 | 2.2×10$^5$ |
| H$^{13}$CO$^+$ | 1-0 | 86.754288 | 4.16 | 0.4 | 2.7×10$^4$ |

The Gaussian fits provided relative coordinates (or offsets), aspect ratios, effective radii ($R_{eff}$), peak flux densities ($S_{peak}$) and total flux densities ($S_{total}$). The aspect ratio is defined as the ratio between de-convolved semi-major size (a) and semi-minor size (b), and $R_{eff}$ is defined as $R_{eff} = \sqrt{ab}$.

For CS, HC$_3$N and H$_{40\alpha}$, the median values for $S_{peak}$ are 65.6, 22.3 and 46.7 Jy beam$^{-1}$ km s$^{-1}$, respectively. The median values for $S_{total}$ of CS, HC$_3$N and H$_{40\alpha}$ lines are 188.9, 49.8 and 53 Jy km s$^{-1}$, respectively. The $R_{eff}$ of CS ranges from 0.04 pc to 1.40 pc with a median value of 0.46 pc. The $R_{eff}$ of HC$_3$N ranges from 0.03 pc to 1.28 pc with a median value of 0.39 pc. The $R_{eff}$ of H$_{40\alpha}$ ranges from 0.08 pc to 0.91 pc with a median value of 0.33 pc. While the sources have a range of sizes in each tracer, the median sizes in all three lines are comparable.

### 3.2 Properties of the Ionized Gas

For the sources with detections of H$_{40\alpha}$, we use the equations in Appendix A to derive the properties of their ionized gas.

The rates of ionizations ($Q_0$) are listed in the eighth column of Table 4. The median, mean and standard deviation of $Q_0$ are $1.7 \times 10^{48}$, $9.1 \times 10^{48}$ and $2.5 \times 10^{49}$ s$^{-1}$, respectively. The median $Q_0$ corresponds to an O8.5 star, and the mean $Q_0$ corresponds to an O6 star.

The electron density (n$_e$), calculated according to equations in Appendix B, is listed in Table 4. The min, max, median, mean and standard deviation of n$_e$ are 543.9, 8968.8, 1465.1, 2190.5 and 1864.6 cm$^{-3}$, respectively. Using n$_e$, we could iterate by adjusting $b_u$ (the departure coefficient). However, values for $b_u$ vary only by about 15% between $n_e = 10^2$ and $10^4$ cm$^{-3}$, so the effect is smaller than other uncertainties.

The masses of ionized gas, computed from equations in Appendix B are listed in Table 4. The min, max, median, mean, and standard deviation are 0.14, 200.74, 4.67, 16.12 and 33.71 M$_\odot$, respectively.

### 3.3 The L$_{bol}$-L$'_{mol}$ scaling relations

The molecular line luminosities (L$'_{mol}$) are computed from the observed line fluxes. For comparison to single-dish work, we provide line luminosities in K km s$^{-1}$ pc$^2$, and their logarithmic values are compiled in Table 5. The line luminosity is calculated by Equation 1 following Solomon et al. (1997):

$$L'_{mol} = 32.5\nu_{obs}^{-2} S_{mol}\Delta\nu D^2, \quad (1)$$

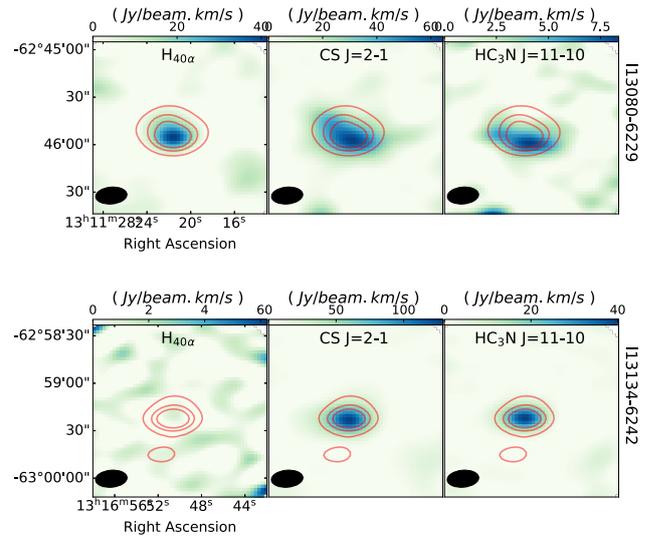

**Figure 1.** Integrated intensity maps of two molecular lines and RRL are shown in color images for two example sources. The 1.3 mm continuum emission is shown in contours. Contours are from 20% to 80% in step of 20% of peak values. The peak values of 1.3 mm continuum emission for I13080-6229 and I13134-6242 are 1.52 and 0.15 Jy beam$^{-1}$, respectively. The black ellipses represent the beams in observations. The integrated intensity maps of the other sources are available on line.

where, $L'_{mol}$ is the the molecular luminosity in K km/s pc$^2$, $S_{mol}\Delta\nu$ is the velocity-integrated flux in Jy km s$^{-1}$, $\nu_{obs}$ is the line frequency in GHz, D is the distance in kpc. The molecular line luminosities are listed in Table 5.

The bolometric luminosity (L$_{bol}$) has been widely used as a tracer of the recent star formation rate (Gao & Solomon 2004; Wu et al. 2005, 2010). L$_{bol}$ values for the ATOMS sources are calculated from integrating the whole SEDs (Faúndez et al. 2004; Urquhart et al. 2018) and listed in the seventh column of Table 5. In left panel of Figure 2, we plot the L$_{bol}$ as a function of the L$'_{CS}$ for the ATOMS measurements (yellow dots) and literature measurements compiled by Wu et al. (2010) (black squares). In this paper, all relations are determined by linear least squares fits in log-log space. Relations with slopes near unity in logarithmic fits are referred to as linear, while those with slopes significantly less than unity are referred to as sub-linear. The fit toward ATOMS data is shown in blue dashed line. The fit to the Wu et al. (2010) data is shown in green dashed line. Both indicate that the underlying relations are close to linear (the slope of the correlation for ATOMS data alone is 0.98 ± 0.07), but the ATOMS data lies mostly above the data from Wu et al. (2010).

**Table 2.** Gas clumps identified in CS J=2-1.

| IRAS | ID | Offset ('', '') | $r_{aspect}$ | $R_{eff}$ (pc) | $S_{peak}$ (Jy beam$^{-1}$km s$^{-1}$) | $S_{total}$ (Jy km s$^{-1}$) |
|---|---|---|---|---|---|---|
| I08076-3556 | 1 | (-10.45,10.51) | 2.13 | 0.04 | 8.83 | 17.96 |
| I08303-4303 | 1 | (-12.53,4.81) | 1.60 | 0.32 | 51.08 | 235.24 |
| I08448-4343 | 1 | (-25.63,11.31) | 1.63 | 0.06 | 28.06 | 44.69 |
|  | 2 | (-4.87,0.50) | 1.63 | 0.08 | 50.29 | 150.51 |
| I08470-4243 | 1 | (-8.87,4.34) | 1.59 | 0.25 | 79.57 | 284.97 |
| I09002-4732 | 1 | (-12.60,-10.59) | 1.22 | 0.08 | 55.45 | 66.30 |
|  | 2 | (22.74,21.83) | 2.85 | 0.13 | 41.40 | 122.80 |
| I09018-4816 | 1 | (-3.39,2.86) | 1.44 | 0.36 | 61.93 | 289.05 |

The full catalogue is available on line.

**Table 3.** Gas clumps identified in HC$_3$N J=11-10.

| IRAS | ID | Offset ('', '') | $r_{aspect}$ | $R_{eff}$ (pc) | $S_{peak}$ (Jy beam$^{-1}$km s$^{-1}$) | $S_{total}$ (Jy km s$^{-1}$) |
|---|---|---|---|---|---|---|
| I08076-3556 | 1 | (-13.23,15.46) | 1.57 | 0.05 | 4.77 | 16.10 |
| I08303-4303 | 1 | (-7.89,6.78) | 1.28 | 0.23 | 23.39 | 59.88 |
| I08448-4343 | 1 | (-30.00,15.63) | 1.38 | 0.07 | 9.27 | 22.64 |
|  | 2 | (-2.87,1.28) | 1.95 | 0.09 | 7.84 | 34.30 |
| I08470-4243 | 1 | (-11.47,2.84) | 1.84 | 0.27 | 12.00 | 54.28 |
| I09002-4732 | 1 | (-12.29,-9.42) | 1.64 | 0.09 | 14.75 | 19.80 |
|  | 2 | (22.95,-11.31) | 1.59 | 0.09 | 8.64 | 13.21 |
|  | 3 | (-11.32,-40.54) | 3.36 | 0.14 | 6.86 | 22.13 |

The full catalogue is available on line.

**Table 4.** Gas clumps identified in H$_{40\alpha}$.

| IRAS | ID | Offset ('', '') | $r_{aspect}$ | $R_{eff}$ (pc) | $S_{peak}$ (Jy beam$^{-1}$km s$^{-1}$) | $S_{total}$ (Jy km s$^{-1}$) | $\log(Q_0)$ (s$^{-1}$) | $\log(n_e)$ (cm$^{-3}$) | $\log(M_{ion})$ (M$_\odot$) |
|---|---|---|---|---|---|---|---|---|---|
| I09002-4732 | 1 | (-0.27,-9.30) | 1.57 | 0.08 | 116.85 | 112.35 | 47.91 | 3.89 | -0.35 |
| I12320-6122 | 1 | (2.45,7.31) | 1.79 | 0.20 | 22.78 | 17.29 | 48.01 | 3.30 | 0.34 |
| I12326-6245 | 1 | (-2.10,0.43) | 1.77 | 0.28 | 93.81 | 75.88 | 48.91 | 3.53 | 1.00 |
| I12383-6128 | 1 | (-3.82,-3.39) | 2.39 | 0.22 | 3.31 | 3.14 | 47.23 | 2.86 | -0.01 |
| I13080-6229 | 1 | (-1.74,-1.62) | 1.37 | 0.33 | 41.98 | 63.72 | 48.67 | 3.31 | 0.98 |
| I13291-6229 | 1 | (9.30,14.82) | 1.48 | 0.21 | 3.47 | 3.52 | 47.17 | 2.86 | -0.06 |
|  | 2 | (-24.63,-6.96) | 2.11 | 0.17 | 3.91 | 2.56 | 47.04 | 2.94 | -0.27 |
| I13291-6249 | 1 | (1.20,2.00) | 1.89 | 0.52 | 16.63 | 15.16 | 48.65 | 3.00 | 1.27 |
| I13471-6120 | 1 | (2.70,5.33) | 1.02 | 0.36 | 38.19 | 32.75 | 48.69 | 3.26 | 1.05 |

The full catalogue is available on line.

This may be due to different source selection in the two samples. In particular, the ATOMS sample has a higher average bolometric luminosity than the sample in Wu et al. (2010). In addition, ACA observations filter out very extended emission and only detect the dense parts of clumps, an effect that leads to a lower molecular line luminosity than in single dish observations.

In Figure 2(b), we plot $L_{bol}$ as a function of the $L'_{HC_3N}$ for the ATOMS measurements (green dots). The relation also has a slope near unity (0.96 ± 0.08). It suggests that HC$_3$N J=11-10 can also trace dense gas and be used for studying star formation scaling relations.

### 3.4 The $L_{bol}$-to-$L'_{mol}$ ratios

$L_{bol}$-to-$L'_{mol}$ ratios is often interpreted in terms of the star formation efficiency (SFE). In this section, we investigate how $L_{bol}$-to-$L'_{mol}$ ratios changes in different Galactic environments and how formed H II regions affect these ratios.

Figure 3 plots the ratio, $L_{bol}$-to-$L'_{mol}$ versus Galactocentric distances ($R_{GC}$, compiled by Liu et al. 2020a). Jiménez-Donaire et al. (2019) found SFE traced by $L_{bol}$-to-$L'_{mol}$ tends to increase with increasing $R_{GC}$ in most of their targeted external galaxies. Eden et al. (2015) also show that ratio of infrared luminosity to the mass of the clump is enhanced in spiral arms compared to the interarm regions when averaged over kiloparsec scales. However, there is no

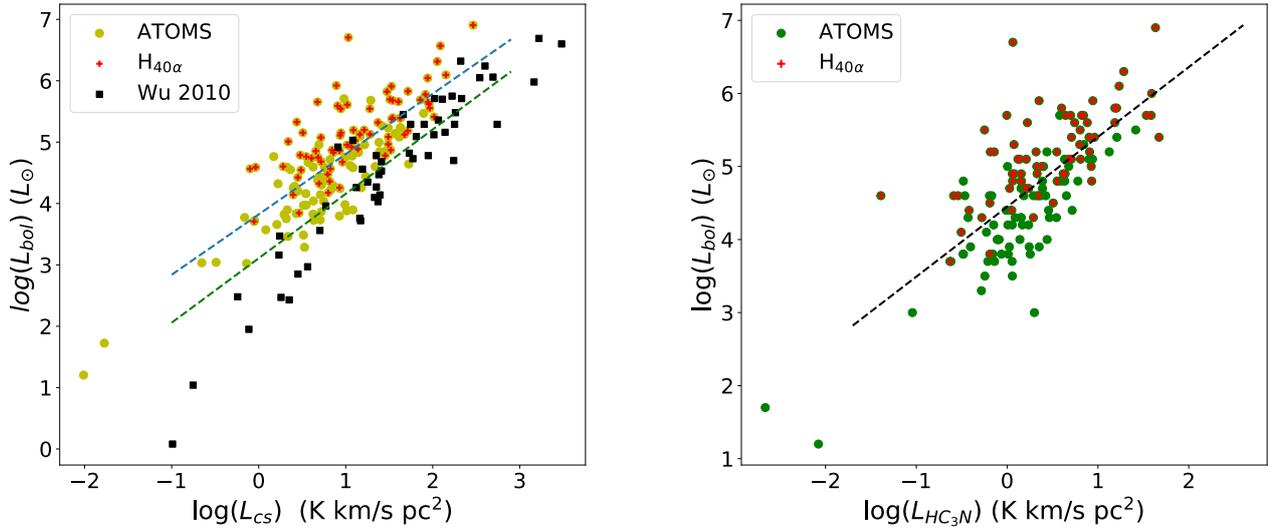

**Figure 2.** $L_{bol}$ - $L'_{mol}$ relations for CS J=2-1 (left panel) and HC$_3$N J=11-10 (right panel). In panel (a), the data for ATOMS are shown as yellow dots. The black cube represent literature measurements compiled by Wu et al. (2010). The linear fit toward ATOMS data is shown in blue dashed line. The slope and offset of blue dashed line are 0.98 ± 0.07 and 3.83 ± 0.08. The linear fit toward Wu et al. (2010) data is shown in a green dashed line. The slope and offset of green dashed line are 1.05 ± 0.08 and 3.11 ± 0.15. In panel (b), the data for ATOMS are shown as green dots. The linear fit toward HC$_3$N data is shown in black dashed line. The slope and offset of black dashed line are 0.96 ± 0.08 and 4.45 ± 0.06. The red crosses in both panels show the ATOMS sources that are detected in H$_{40\alpha}$ emission.

**Table 5.** Luminosities of sources.

| ID | IRAS | RA | DEC | Distance (kpc) | $R_{GC}$ (kpc) | log($L_{bol}$) ($L_\odot$) | log($L'_{CS}$) (K km/s pc$^2$) | log($L'_{HC_3N}$) (K km/s pc$^2$) |
|---|---|---|---|---|---|---|---|---|
| 1 | I08076-3556 | 08:09:32.39 | -36:05:13.2 | 0.4 | 8.5 | 1.20 | -2.01 | -2.08 |
| 2 | I08303-4303 | 08:32:08.34 | -43:13:54.0 | 2.3 | 9.0 | 3.83 | 0.62 | 0.01 |
| 3 | I08448-4343 | 08:46:32.90 | -43:54:35.9 | 0.7 | 8.4 | 3.04 | -0.49 | -1.04 |
| 4 | I08470-4243 | 08:48:47.07 | -42:54:31.0 | 2.1 | 8.8 | 4.04 | 0.63 | -0.11 |
| 5 | I09002-4732 | 09:01:54.24 | -47:44:00.8 | 1.2 | 8.4 | 4.59 | -0.04 | -0.59 |
| 6 | I09018-4816 | 09:03:32.84 | -48:28:10.0 | 2.6 | 8.8 | 4.72 | 0.82 | 0.39 |
| 7 | I09094-4803 | 09:11:07.29 | -48:15:48.7 | 9.6 | 12.7 | 4.60 | 1.34 | -0.19 |
| 8 | I10365-5803 | 10:38:32.46 | -58:19:05.9 | 2.4 | 8.0 | 4.28 | 0.41 | 0.09 |
| 9 | I11298-6155 | 11:32:05.70 | -62:12:24.3 | 10 | 10.1 | 5.23 | 1.49 | 0.44 |

The full catalogue is available on line.

clear trend for $L_{bol}$-to-$L'_{mol}$ ratios against $R_{GC}$ in our study. Liu et al. (2020b) also showed similar results with J = 1-0 transitions of H$^{13}$CN, H$^{13}$CO$^+$, HCN and HCO$^+$. We also show these ratios for two groups (with and without H$_{40\alpha}$ emission or H II regions) separately. Again, there is no clear trend for $L_{bol}$-to-$L'_{mol}$ ratios against $R_{GC}$ whether or not UC H II regions have formed. A number of other studies (Eden et al. 2013; Ragan et al. 2018; Urquhart et al. 2021) also found that the distribution of the dense gas fraction in Galactic giant molecular clouds is relatively flat across the disc and there is no evidence of any significant enhancements that might be attributable to spiral arms.

We present the log($L_{bol}$-to-$L'_{mol}$) distributions of different molecular line tracers in Figure 4 for two groups of sources, i.e., those with and without H$_{40\alpha}$ emission (or H II regions). Line luminosities for the J = 1-0 transitions of H$^{13}$CN, H$^{13}$CO$^+$, HCN and HCO$^+$ were compiled in Liu et al. (2020b). We test whether the log($L_{bol}$-to-$L'_{mol}$) values for clumps with H II regions or without H II regions could be drawn from the same distribution with Kolmogorov–Sminov test. The p-values from Kolmogorov–Sminov tests are summerized in the last column of Table 6. All the p-value for different molecular line tracers are near to zero. This implies that $L_{bol}$-to-$L'_{mol}$ ratios (or SFE) for sources with and without H$_{40\alpha}$ emission are significantly different. In other words, clumps associated with H II regions show significantly larger $L_{bol}$-to-$L'_{mol}$ ratios.

We list the min, max, mean, median, standard deviation (std) of log($L_{bol}$-to-$L'_{mol}$) for different tracers in Table 6. For every tracer, the first row shows the parameters for all the ATOMS sources. The second row shows data for sources with H$_{40\alpha}$ emission and the third row for sources without H$_{40\alpha}$ emission. The mean values of log($L_{bol}$-to-$L'_{mol}$) ratios between two samples offset by ∼0.5 dex. In fact, this offset is remarkably constant over all tracers with a mean value of 0.52 ± 0.04. We will discuss physical explanations for these different distributions for sources with and without $H_{40\alpha}$ emission in Section 4.

## 4 DISCUSSION

### 4.1 Star formation rates

Star formation rates in other galaxies are routinely estimated from a variety of indicators, including the bolometric luminosity and tracers of ionization by massive stars, such as recombination lines. These methods rely on models of star formation that extend over a long period and fully sample the initial mass function. Application of these models to individual regions of star formation in the Milky Way is not rigorously justified, but agreement between tracers with different dependence on the duration and IMF sampling provide empirical support for their use (e.g., Vutisalchavakul et al. 2016).

Using $L_{bol}$ values listed in the seventh column of Table 5, following Kennicutt & Evans (2012) but converting from $M_\odot$ yr$^{-1}$ to $M_\odot$ Myr$^{-1}$, we can relate $L_{bol}$ to SFR as:

$$\log(\frac{SFR}{M_\odot \text{ Myr}^{-1}}) = \log(\frac{L_{bol}}{\text{ergs s}^{-1}}) - 37.41 \qquad (2)$$

The star formation rates (SFR$_{bol}$) derived from the $L_{bol}$ are listed in the fifth column of Table 7.

Millimeter or submillimter RRLs provide a relatively new way to estimate star formation rates (SFRs), so far mostly in external galaxies (Yun et al. 2004; Scoville & Murchikova 2013; Bendo et al. 2016, 2017; Michiyama et al. 2020). Here we test their use in Galactic regions by comparing with SFRs estimated from bolometric luminosities. For the 75 sources with $H_{40\alpha}$ line detections, we estimate their SFRs with $H_{40\alpha}$ integrated line flux density ($S\Delta v$) using the equations in the Appendix A. These SFR$_{H_{40\alpha}}$ are given in Table 7.

Vutisalchavakul et al. (2016) found a slightly sub-linear correlation (slope 0.8) with SFR of radio continuum data and bolometric luminosity. The slightly sub-linear correlation can result from the fact that the radio continuum emission is more sensitive to the upper end of the IMF, as can be seen from Kennicutt & Evans (2012) and Vutisalchavakul et al. (2016). $H_{40\alpha}$ emission also requires stars producing EUV photons with energy above 13.6 eV. Kim et al. (2017) has found a strong correlation between the integrated 6 cm radio continuum and millimeter RRL emission in Galactic HII regions, indicating that $H_{40\alpha}$ emission could also well trace the upper end of the IMF.

In addition, $L_{bol}$ also begins to underestimate the SFRs for sources with SFRs less than about 5 $M_\odot$ Myr$^{-1}$, corresponding to a total far-infrared luminosity of $10^{4.5}$ $L_\odot$ (Wu et al. 2005; Vutisalchavakul & Evans 2013; Vutisalchavakul et al. 2016). Wu et al. (2005) found a rapid drop in SFR below $L_{bol} = 10^{4.5}$ $L_\odot$, due to poor sampling of the upper part of the IMF. Therefore, we use only sources with SFR$_{H_{40\alpha}} \geqslant$ 5 $M_\odot$ Myr$^{-1}$ and SFR$_{bol} \geqslant$ 5 $M_\odot$ Myr$^{-1}$ for comparison. We plot the ratio SFR$_{H_{40\alpha}}$/SFR$_{bol}$ versus SFR$_{bol}$ in Figure 5. The mean, median and std of the log values of SFR$_{H_{40\alpha}}$/SFR$_{bol}$ ratio are $-0.01$, 0.02 and 0.29, respectively. While both estimates of SFR are uncertain, the excellent agreement suggests that either can provide an estimate of the SFR. There is no obvious dependence of the ratio on SFR$_{bol}$ for values above 5 $M_\odot$ Myr$^{-1}$.

### 4.2 The differences between sources with and without $H_{40\alpha}$ emission

Figure 4 shows that $L_{bol}$-to-$L'_{mol}$ ratios for sources with and without $H_{40\alpha}$ emission are significantly different, indicating that the star formation efficiency (SFE) of these clumps may be very different. Recently, Elia et al. (2021) found large difference in the $L_{bol}$-to-mass ratios (also tracing SFEs) between HII region candidates and protostellar objects in Hi-GAL compact sources. The masses of Hi-GAL compact sources are estimated from dust continuum emission. They find that the median value of $\log(L_{bol}/\text{mass})$ for HII region candidates is about 20 times larger than that for protostellar objects. This factor is much larger than what we find here for ATOMS sources. This is because they include many more evolved HII regions, some of which may have broken out of the dense clump. In contrast, the HII region candidates in the ATOMS sample are mostly in the ultra-compact HII region phase.

Before exploring the evolutionary explanation, we consider an alternative explanation. Perhaps a large amount of gas has been ionized or dispersed by formed OB stars (HII regions) and this gas is no longer counted in the molecular emission. We can investigate how much gas has been ionized in the ATOMS sources. The mass of ionized gas in an HII region with 27% helium can be estimated with total flux of $H_{40\alpha}$ by equations in Appendix B. The mass of ionized gas is summarized in Table 4 and Table 7. For the 4 sources with multiple objects, we add $M_{ion}$ to get the total ionized gas masses. Then we derive the ionized gas mass to molecular gas mass ratios. Because neither the molecular lines nor 3 mm continuum emission in the ATOMS data are well calibrated to determine mass, we use the masses of molecular gas ($M_{mol}$ or $M_{clump}$) compiled by Liu et al. (2020a), which are estimated from dust continuum observations with single dishes. The masses of molecular gas ($M_{mol}$ or $M_{clump}$) are shown in the 7th column of Table 7. Left panel in figure 6 shows the distribution of $\log(M_{ion}/M_{mol})$. The min, max, mean, median and std of the log values of $M_{ion}/M_{mol}$ are -4.56, -1.90, -2.70, -2.66 and 0.47, respectively. The median mass of ionized gas is a factor of ∼500 smaller than the median molecular mass, indicating that ionized gas is negligible for the ATOMS sources.

The next step is to estimate the gas depletion time $\tau_{dep}$=Mass/SFR. We investigate the difference in $\tau_{dep}$ between sources with and without HII regions. Figure 7 shows the correlation between total gas mass of clumps and $\tau_{dep}$. We only focus on sources with SFR$_{H_{40\alpha}} \geqslant$ 5 $M_\odot$ Myr$^{-1}$ or SFR$_{bol} \geqslant$ 5 $M_\odot$ Myr$^{-1}$ to avoid the underestimations of SFRs. Middle and right panels in figure 6 show the distribution of $\log(\tau_{dep})$ for sources with and without $H_{40\alpha}$ emission. The median, mean and std of $\log(\tau_{dep}/\text{Myr})$ of all sources are 2.09, 2.15 and 0.35, respectively. For sources associated with HII regions, the median, mean and std of $\log(\tau_{dep}/\text{Myr})$ are 1.98, 2.05 and 0.32, respectively. For sources without HII regions, the median, mean and std of $\log(\tau_{dep}/\text{Myr})$ are 2.26, 2.31 and 0.33, respectively. Thus depletion times are shorter by about 0.25 dex for sources with HII regions. Wu et al. (2010) also found a decreasing trend in $\tau_{dep}$ (or increasing trend in SFE) from clouds without HII regions to those with UC HII regions to those with compact HII regions to those with larger HII regions.

The shorter $\tau_{dep}$ or higher SFE for sources with $H_{40\alpha}$ emission is not caused simply by turning molecular gas into ionized gas. The SFE may be higher in clumps with $H_{40\alpha}$ because the IMF is better sampled and OB stars are needed for ionization. The argument against this is that $H_{40\alpha}$ and $L_{bol}$ predict similar SFR over the whole range of SFR as long as SFR $\geqslant$ 5 $M_\odot$ Myr$^{-1}$. The most likely explanation is that the SFE is higher in regions with $H_{40\alpha}$ emission simply because star formation has proceeded further from massive, but not yet ionizing, protostars to the formation of OB stars. In other word, this is an evolutionary effect.

**Table 6.** Statistics of log(L$_{bol}$/L$'_{mol}$) for different tracers. For every tracer, the first row shows the parameters for all the log(L$_{bol}$/L$'_{mol}$) data. The second row for data with H$_{40\alpha}$ emission and the third row for data without H$_{40\alpha}$ emission.

| Molecule | H$_{40\alpha}$ | min | max | mean | median | std | p-value |
|---|---|---|---|---|---|---|---|
| CS(2-1) |  | 2.77 | 5.67 | 3.80 | 3.76 | 0.49 | 1.86e-8 |
|  | Y | 3.34 | 5.67 | 4.05 | 3.98 | 0.45 |  |
|  | N | 2.77 | 4.72 | 3.56 | 3.53 | 0.41 |  |
| HC$_3$N(11-10) |  | 2.70 | 6.64 | 4.44 | 4.36 | 0.55 | 1.10e-10 |
|  | Y | 3.73 | 6.64 | 4.73 | 4.67 | 0.50 |  |
|  | N | 2.70 | 5.28 | 4.15 | 4.17 | 0.43 |  |
| H$^{13}$CN(1-0) |  | 3.43 | 6.63 | 4.38 | 4.32 | 0.50 | 2.97e-7 |
|  | Y | 3.87 | 6.63 | 4.64 | 4.54 | 0.49 |  |
|  | N | 3.43 | 5.30 | 4.16 | 4.16 | 0.39 |  |
| H$^{13}$CO$^+$(1-0) |  | 3.26 | 6.53 | 4.59 | 4.57 | 0.52 | 9.01e-9 |
|  | Y | 4.03 | 6.53 | 4.87 | 4.84 | 0.49 |  |
|  | N | 3.26 | 5.36 | 4.35 | 4.30 | 0.41 |  |
| HCN(1-0) |  | 2.47 | 5.53 | 3.67 | 3.63 | 0.53 | 3.81e-7 |
|  | Y | 3.12 | 5.53 | 3.96 | 3.98 | 0.47 |  |
|  | N | 2.47 | 4.55 | 3.43 | 3.39 | 0.45 |  |
| HCO$^+$(1-0) |  | 2.43 | 5.57 | 3.86 | 3.80 | 0.53 | 5.95e-5 |
|  | Y | 3.30 | 5.57 | 4.12 | 4.08 | 0.48 |  |
|  | N | 2.43 | 4.72 | 3.63 | 3.62 | 0.46 |  |

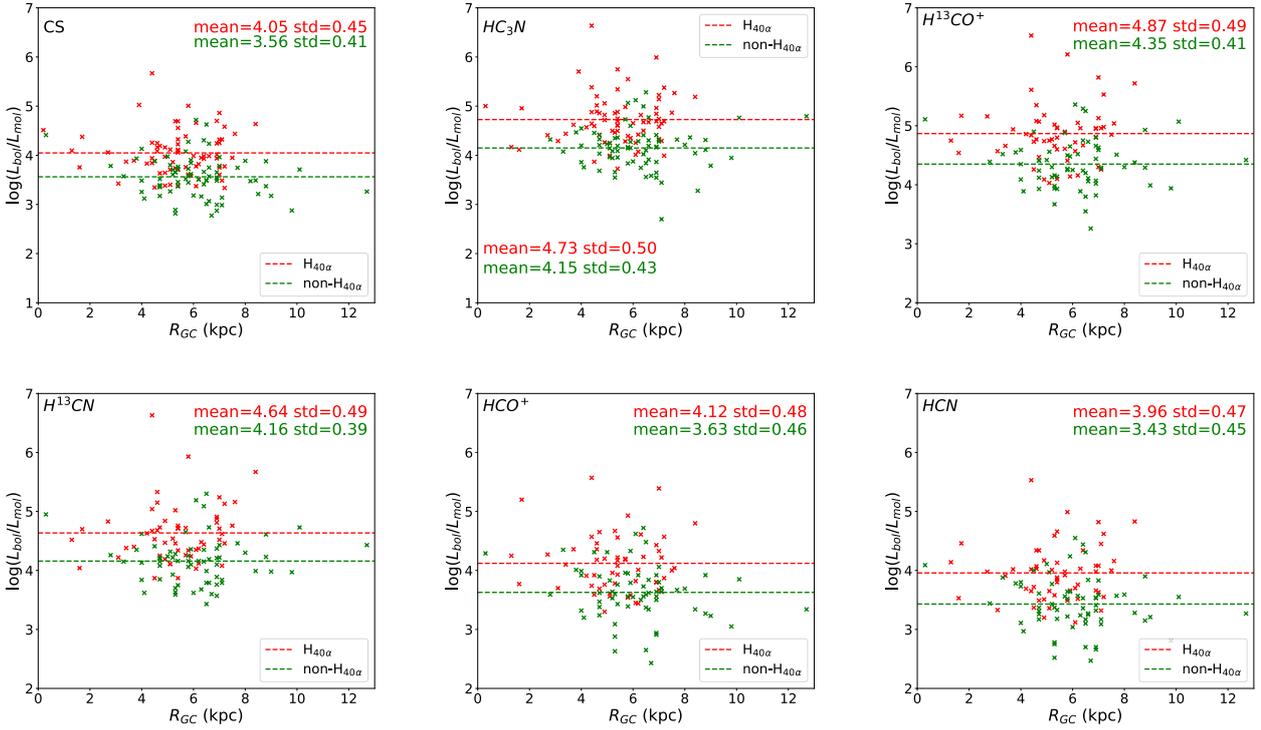

**Figure 3.** L$_{bol}$-to-L$_{mol}$ ratios as a function of galactocentric distances (R$_{GC}$). The red and green dots represent sources with (red) or without (green) H$_{40\alpha}$ emission.

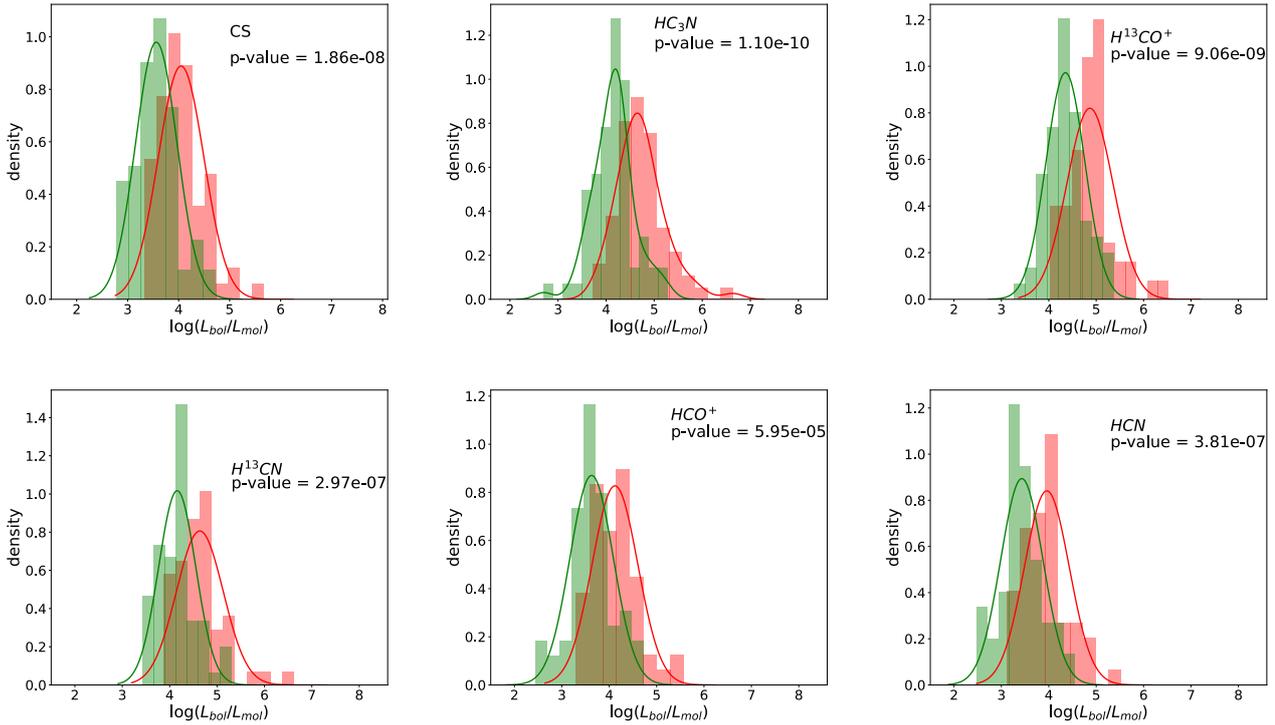

**Figure 4.** Density distributions of $L_{bol}$-to-$L'_{mol}$ ratios. The red and green histograms represent sources with $H_{40\alpha}$ emission and without $H_{40\alpha}$ emission, respectively.

**Table 7.** SFRs and gas masses of ATOMS sources.

| ID | IRAS | $R_{eff}$ (pc) | $\log(L_{H_{40\alpha}})$ (Jy km/s kpc$^2$) | $\log(SFR_{bol})$ ($M_\odot$ Myr$^{-1}$) | $\log(SFR_{H_{40\alpha}})$ ($M_\odot$ Myr$^{-1}$) | $\log(M_{clump})$ ($M_\odot$) | $\log(M_{ion})$ ($M_\odot$) |
|---|---|---|---|---|---|---|---|
| 1 | I08076-3556 | 0.04 | - | -2.62 | - | 0.7 | - |
| 2 | I08303-4303 | 0.16 | - | 0.00 | - | 2.4 | - |
| 3 | I08448-4343 | 0.07 | - | -0.78 | - | 1.6 | - |
| 4 | I08470-4243 | 0.16 | - | 0.22 | - | 2.4 | - |
| 5 | I09002-4732 | 0.12 | 3.31 | 0.77 | 0.79 | 2.4 | -0.35 |
| 6 | I09018-4816 | 0.22 | - | 0.89 | - | 3.0 | - |
| 7 | I09094-4803 | 0.70 | - | 0.78 | - | 3.1 | - |
| 8 | I10365-5803 | 0.22 | - | 0.45 | - | 2.7 | - |
| 9 | I11298-6155 | 0.68 | - | 1.41 | - | 3.4 | - |

The full catalogue is available on line.

## 5 CONCLUSION

In this paper, we have studied 146 ATOMS sources using CS J=2-1, HC$_3$N J=11-10 and H$_{40\alpha}$. We found empirical relations between $L_{bol}$ and $L'_{mol}$. We calculate SFR with H$_{40\alpha}$, compare relations between SFR$_{L_{bol}}$ and SFR$_{H_{40\alpha}}$, and derive gas depletion times. Our main results are summarized as follows:

(i) We extracted 177, 185 and 75 compact sources from CS J=2-1, HC$_3$N J=11-10 and H$_{40\alpha}$, respectively.

(ii) The correlations between $L_{bol}$, tracing the SFR, and molecular line luminosities $L'_{mol}$ of the CS J=2-1 and HC$_3$N J=11-10 appear linear. Both of these two lines can be used to trace dense gas.

(iii) The $L_{bol}$-to-$L'_{mol}$ ratios stays constant against $R_{GC}$ suggesting that SFE at the clump scale does not depend on $R_{GC}$.

(iv) The star forming clumps with H$_{40\alpha}$ emission (thus HII regions) show higher $L_{bol}$-to-$L'_{mol}$ ratios (or SFE) than those without H$_{40\alpha}$ emission. The mean values of $\log(L_{bol}$-to-$L'_{mol})$ ratios between the two sub-samples are offset by ∼0.5 dex.

(v) The SFRs calculated from H$_{40\alpha}$ RRL and $L_{bol}$ are very similar to each other on average, encouraging their use.

(vi) The mass of ionized gas is a factor of ∼500 below the molecular mass, indicating that the higher star formation rate per unit mass for regions associated with HII regions is not caused simply by ionizing more gas.

(vii) A plausible explanation for the higher SFEs (or shorter $\tau_{dep}$) for clumps associated with HII regions is that the SFE is higher simply because the region is more evolved.

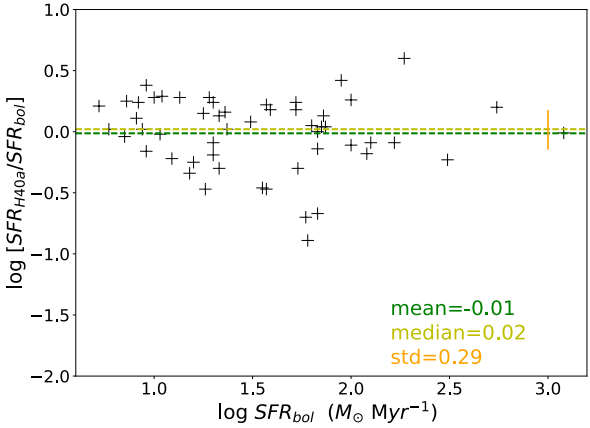

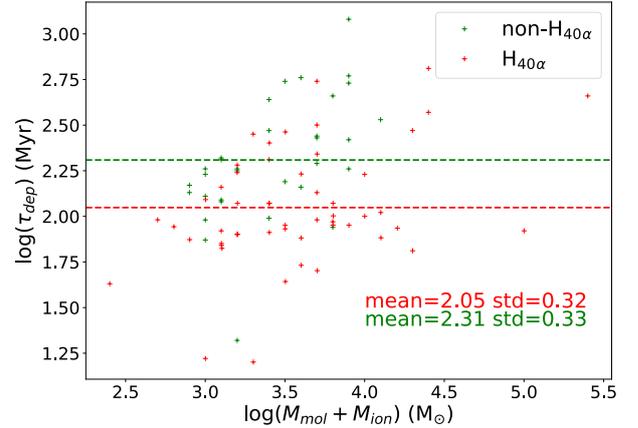

**Figure 5.** The logarithm of the ratio, $SFR_{H40\alpha}/SFR_{bol}$, is plotted versus the logarithm of $SFR_{bol}$. The green, yellow and orange lines represent the log of mean , median and std of $SFR_{H40\alpha}/SFR_{bol}$. We use only sources with $SFR_{H40\alpha} \geqslant 5$ $M_\odot$ $Myr^{-1}$ and $SFR_{bol} \geqslant 5$ $M_\odot$ $Myr^{-1}$ for comparison.

**Figure 7.** The correlation between $\tau_{dep}$ and mass of molecular cloud add mass of $H_{40\alpha}$ emission area. The red and green crosses represent sources with $H_{40\alpha}$ emission and without $H_{40\alpha}$ emission, respectively. We use only sources with $SFR_{H40\alpha} \geqslant 5$ $M_\odot$ $Myr^{-1}$ and $SFR_{bol} \geqslant 5$ $M_\odot$ $Myr^{-1}$ for comparison.

## DATA AVAILABILITY

The data underlying this article are available in the article and in its online supplementary material.

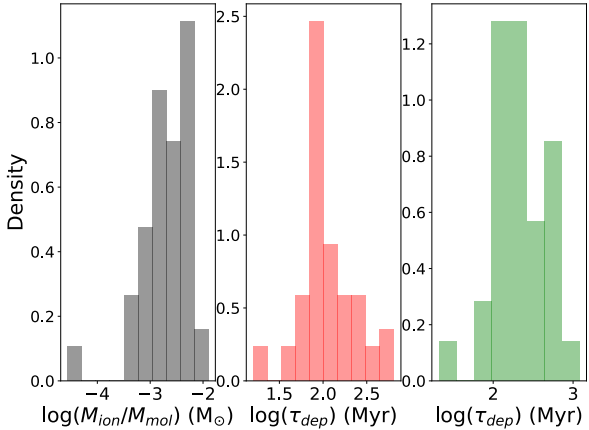

**Figure 6.** The Distribution of $M_{ion}/M_{mol}$ and $\tau_{dep}$ from left to right. The red represents clumps with $H_{40\alpha}$ and green represents the clumps without $H_{40\alpha}$ emission, respectively. For the middle and right panels, we use only sources with $SFR_{H40\alpha} \geqslant 5$ $M_\odot$ $Myr^{-1}$ and $SFR_{bol} \geqslant 5$ $M_\odot$ $Myr^{-1}$.

## ACKNOWLEDGEMENTS

This work was partially supported by National Natural Science Foundation of China (NSFC) through grants No. 11988101, No.12073061, and No.12122307. NJE thanks the Department of Astronomy at the University of Texas at Austin for research support. Tie Liu acknowledges the supports by the international partnership program of Chinese academy of sciences through grant No.114231KYSB20200009, and Shanghai Pujiang Program 20PJ1415500. C.W.L. is supported by Basic Science Research Program through the National Research Foundation of Korea (NRF)funded by the Ministry of Education, Science and Technology(NRF-2019R1A2C1010851). J.W. acknowledge support from National Key RD Program of China grant No.2017YFA0402600.

## APPENDIX A: FROM RADIO RECOMBINATION LINES TO STAR FORMATION RATES

To derive star formation rates (SFR) from a Radio Recombination Line (RRL), the following steps are followed. For each step, we note assumptions and uncertainties.

(i) The total flux density ($S\Delta v$) and effective radius $R_{\rm eff}$ are derived from an elliptical Gaussian fit to the integrated intensity image of $H_{40\alpha}$ emission. The linewidth is checked for consistency with expectations for an Hɪɪ region. In ATOMS Paper III (Liu et al. 2021), we noted that the linewidth of $H_{40\alpha}$ line increases with decreasing size, suggesting that pressure broadening or dynamical motions could be important in the most compact sources. However the linewidths were generally consistent with primarily thermal broadening in Hɪɪ regions, and there is no evidence in the linewidths of strong maser emission.

(ii) The line luminosity is computed from the total flux density and the distance:

$$L(H_{40\alpha}) = 4\pi D^2 S\Delta v \quad (A1)$$

Uncertainties in distance can introduce major uncertainties because $L \propto D^2$. The distances to sources were taken from Paper I of this series (Liu et al. 2020a). In Paper III (Liu et al. 2021), We recalculated distances with new kinematic models and found only minor differences, but were able to assess the median uncertainty in distance to be 7%.

(iii) The total rate of recombinations is computed from $R_{\rm rec} = n_e n_p V \alpha_B$, where $V$ is the volume and $\alpha_B$ is the Case B total recombination coefficient. We neglect ionizations of helium at this point. The line luminosity (in frequency units) is $L_\nu(H_{40\alpha}) = n_e n_p V \epsilon$, where $L_\nu(H_{40\alpha}) = L(H_{40\alpha})\nu/c$ and $\epsilon$ is the efficiency for producing $H_{40\alpha}$ photons per recombination. So, we have

$$R_{\rm rec} = L_\nu(H_{40\alpha}) \frac{\alpha_B}{\epsilon}, \quad (A2)$$

At $T_e = 10^4$ K, $\alpha_B = 2.54 \times 10^{-13}$ and the dependence on temperature is rather weak (equation 14.6 of 2011piim.book.....D). We assume $T_e = 10^4$ K. The efficiency factor is

$$\epsilon = b_u f_u(LTE) A_{ul} h\nu_{ul} \quad (A3)$$

where $f_u(LTE)$ is the fractional population of level $u$ in LTE, $b_u$ is the departure coefficient, $A_{ul}$ is the spontaneous emission coefficient, and $\nu_{ul}$ is the frequency of the transition. The value of $\epsilon$ comes from calculations of excitation including recombination, radiative, and collisional processes. We assume that any maser effects are weak, so the uncertainties are less than those introduced by the distance uncertainties. Ideally, we would check other RRLs for signs of anomalous excitation, but lacking those, we rely on theory. From various sources, we determine that $b_u = 0.7866$ (Zhu et al. 2019), $f_u(LTE) = 7.03 \times 10^{-19}$ (Peters et al. 2012), $A_{ul} = 54.7973$ s$^{-1}$ (Brocklehurst 1971), and $\nu_{ul} = 99.023 \times 10^9$ Hz (Liu et al. 2020a), yielding $\epsilon = 1.99 \times 10^{-32}$. $f_u(LTE)$ depends on the electron temperature, $T_e$ and $b_u$ depends on both $T_e$ and $n_e$. We assume $T_e = 10^4$ K and $n_e = 10^4$ cm$^{-3}$. From figure 1 of Zhu et al. (2019), we estimate variations of about 10% for densities as low as 100 cm$^{-3}$ and temperatures as low as 5000 K. If the line is masing, the value of $\epsilon$ could be much higher. We think this is unlikely to be a major effect for this transition, but further calculations are underway (J-Z Wang, personal communication).

(iv) The production rate of ionizing photons is derived from the $R_{\rm rec}$ by assuming that the rate of ionizations ($Q_0$) equals the rate of recombinations, ignoring helium ionizations. This is the usual assumption because steady state is achieved rapidly, with a recombination timescale of $t_{\rm rec} = 1.22 \times 10^3/n_2$ yr, where $n_2$ is the density in units of 100 cm$^{-3}$. This is much shorter than the dynamical timescale for expansion, $t_{\rm dyn} = R_S/c_s = 2.39 \times 10^5 \times Q_{0,49}^{1/3}/n_2^{2/3}$ yr, where $Q_{0,49}$ is the production rate of ionizing photons in units of $1 \times 10^{49}$ s$^{-1}$, corresponding to an O6V star. These are equations 15.7 and 15.8 from Draine (2011). For the mean values of $n_e$ and $Q_0$, we estimate $t_{\rm rec} = 6$ yr and $t_{\rm dyn} = 1.9 \times 10^6$ yr, so the assumption of steady state seems safe. However, these textbook results can be wrong in extreme circumstances, especially at very early times or in very inhomogenous environments. Another issue is that some of the ionizing photons may be absorbed by dust. Binder & Povich (2018b) found that about 34% of the ionizing photons were absorbed by dust in their study of 28 massive Galactic regions. Both these effects would cause an underestimate in the ionization rate.

(v) The production rate of ionizing photons is related to the presence of massive stars. Assuming a well-sampled IMF and a timescale

long enough that the birth and death of massive stars have reached steady state, the ionization rate can be related to the SFR. Based on Starburst99 (Leitherer et al. 1999) models by Binder & Povich (2018b), using a IMF model from Kroupa & Weidner (2003), we assume

$$\frac{SFR(H_{40\alpha})}{M_\odot\,\mathrm{Myr}^{-1}} = 7.5\times 10^{-48} Q_0 = 7.5\times 10^{-48} L_\nu(H_{40\alpha})\frac{\alpha_B}{\epsilon} \quad \mathrm{(A4)}$$

For the values of frequency and $\epsilon$ discussed above, and converting to observational units of Jy km s$^{-1}$kpc$^2$ for $L(H_{40\alpha})$, we have

$$Q_0 = 4.010\times 10^{44}\left[\frac{L(H_{40\alpha})}{\mathrm{Jy\ km/s\ kpc}^2}\right] \quad \mathrm{(A5)}$$

and

$$\frac{SFR(H_{40\alpha})}{M_\odot\,\mathrm{Myr}^{-1}} = 3.008\times 10^{-3}\left[\frac{L(H_{40\alpha})}{\mathrm{Jy\ km/s\ kpc}^2}\right] \quad \mathrm{(A6)}$$

where $L(H_{40\alpha}) = 4\pi(\frac{D}{\mathrm{kpc}})^2(\frac{S\Delta v}{\mathrm{Jy\ km/s}})$. In general, we convert from $L_\nu(H_{40\alpha})$ in cgs units to $L(H_{40\alpha})$ in observer units (Jy km s$^{-1}$kpc$^2$) via

$$L_\nu(H_{40\alpha}) = 3.174\times 10^{23}\nu_{\mathrm{GHz}}L(H_{40\alpha}) = 3.143\times 10^{25}L(H_{40\alpha}). \quad \mathrm{(A7)}$$

While Binder & Povich (2018b) showed that the star formation rates from their relation agreed well with other measures, they cautioned that they are likely to be underestimates because of dust absorption and young ages, especially for compact sources such as ours. Our only justification for using either the FIR or the RRL for star formation rates is that the SFR from the RRL agrees reasonably well with the SFR from the FIR. Both are likely to be underestimates.

## APPENDIX B: DERIVATION OF ELECTRON DENSITY AND IONIZED GAS MASS

The electron density is readily obtained from the observations under the assumptions contained in steps 1 through 3 above.

$$n_e = \left[\frac{R_{\mathrm{rec}}}{\alpha_B V}\right]^{0.5} = \left[\frac{L_\nu(H_{40\alpha})}{\epsilon V}\right]^{0.5} = 3.582\left[\frac{L(H_{40\alpha})}{\mathrm{Jy\ km/s\ kpc}^2}\right]^{0.5}\left[\frac{R_{\mathrm{eff}}}{\mathrm{pc}}\right]^{-1.5} \quad \mathrm{(B1)}$$

where the result is in cm$^{-3}$.

The mass of ionized gas is then computed from

$$M_{\mathrm{ion}} = n_p m_H \mu V = 0.471 M_\odot\left[\frac{L(H_{40\alpha})}{\mathrm{Jy\ km/s\ kpc}^2}\right]^{0.5}\left[\frac{R_{\mathrm{eff}}}{\mathrm{pc}}\right]^{1.5} \quad \mathrm{(B2)}$$

where $\mu$ is the mean mass per proton (1.27 for 27% helium) and we take $n_p = n_e$, neglecting ionization of helium.

This paper has been typeset from a T$_{\mathrm{E}}$X/L$^A$T$_{\mathrm{E}}$X file prepared by the author.